\UseRawInputEncoding
\documentclass[prb,reprint,amsmath,amssymb,showpacs,superscriptaddress,floatfix,longbibliography]{revtex4-1}
\usepackage[breaklinks=true,colorlinks,citecolor=blue,linkcolor=blue,urlcolor=blue]{hyperref}
\usepackage[T1]{fontenc}
\usepackage{epsfig,mathrsfs,color,latexsym,subfigure,marginnote,gensymb,}
\usepackage{graphicx}
\usepackage{makecell}
\usepackage{textalpha}
\usepackage{xr-hyper}
\usepackage{ulem}
\usepackage{hyperref}
\renewcommand{\BibitemShut}[1]{}
\makeatletter
\newcommand*{\addFileDependency}[1]{
  \typeout{(#1)}
  \@addtofilelist{#1}
  \IfFileExists{#1}{}{\typeout{No file #1.}}
}
\makeatother

\newcommand*{\myexternaldocument}[1]{%
    \externaldocument{#1}%
    \addFileDependency{#1.tex}%
    \addFileDependency{#1.aux}%
}

\myexternaldocument{Supplementary}

\begin{document}
\title{Role of interface morphology on the martensitic transformation in pure Fe}
\author{Pawan Kumar Tripathi}
\affiliation{Department of Materials Science and Engineering, Indian Institute of Technology Kanpur, Kanpur 208016, India}
\affiliation{International College of Semiconductor Technology, National Chiao Tung University, Taiwan}
\author{Shivraj Karewar}
\affiliation{Department of Mechanical Engineering, Eindhoven University of Technology, Eindhoven, The Netherlands}
\author{Yu-Chieh Lo}
\affiliation{Department of Material Science and Engineering, National Chiao Tung University, Taiwan}
\author{Somnath Bhowmick}
\email{bsomnath@iitk.ac.in}
\affiliation{Department of Materials Science and Engineering, Indian Institute of Technology Kanpur, Kanpur 208016, India}
\date{\today}

\begin{abstract}
Using classical molecular dynamics simulations, we study austenite to ferrite phase transformation in iron, focusing on the role of interface morphology. We compare two different morphologies; a \textit{flat} interface in which the two phases are joined according to  Nishiyama-Wasserman orientation relationship vs. a \textit{ledged} one, having steps similar to the vicinal surface. We identify the atomic displacements along a misfit dislocation network at the interface leading to the phase transformation. In case of \textit{ledged} interface, stacking faults are nucleated at the steps, which hinder the interface motion, leading to a lower mobility of the inter-phase boundary, than that of flat interface. Interestingly, we also find the temperature dependence of the interface mobility to show opposite trends in case of \textit{flat} vs. \textit{ledged} boundary. We believe that our study is going to present a unified and comprehensive view of martensitic transformation in iron with different interface morphology, which is lacking at present, as \textit{flat} and \textit{ledged} interfaces are treated separately in the existing literature. 
\end{abstract}


%
%
\maketitle
\section{Introduction}
\label{intro}

High temperature Austenite ($\gamma$-phase, face centered cubic or FCC) to low temperature ferrite ($\alpha$-phase, body centered cubic or BCC) phase transformation in iron is crucial, as it governs microstructure, and subsequently various material properties of different types of steels.\cite{moritani, CABALLERO2004251,RAABE20136132,WANG2014268,TOJI2015137} As iron is quenched, the transformation from the high to low temperature phase is known to take place via a diffusionless process, which can further be subdivided in two categories, martensitic and massive. In case of martensitic transformation, a definite orientation relationship (OR) between the parent and product phase is required to facilitate coordinated movement of atoms (at the velocity of the sound \cite{bunshah}), giving rise to the name military transformation. Massive transformation, on the other hand, is civilian in nature, as the atoms move individually and there is no restriction on OR between the parent and product phase. During the transformation, the motion of the interface is dependent on temperature and cooling rate; and it is also affected by various factors like interface morphology, grain boundaries, alloying elements, and precipitates etc.\cite{moritani, CABALLERO2004251,RAABE20136132,WANG2014268,TOJI2015137}

During the phase transition certain transformation paths are followed, depending on the ORs between the parent and product phase. In 1930, Kurdjumov - Sachs (KS) \cite{Kurdjumow1930} identified an OR in mild steel using X-ray diffraction as $\{111\}_{fcc}  ||  \{011\}_{bcc}; [\bar{1}01]_{fcc}  ||  [\bar1\bar11]_{bcc}$ and later, Nishiyama -Wasserman (NW) \cite{nw1934} found a slightly distinct OR as $\{111\}_{fcc}  ||  \{110\}_{bcc} ; [\bar{1}\bar{1}2]_{fcc}  ||   [0\bar{1}1]_{bcc}$ in Fe-30$\%$Ni alloys. Based on the crystal symmetry of the two parent phases, 24 and 12 orientational variants are possible in KS\cite{Kurdjumow1930} and NW\cite{nw1934} ORs, respectively. Several other ORs, like  Bain\cite{bain}, Pitsch\cite{PITSCH1962897}  and Greninger and Troiano\cite{gt} have also been discussed in the literature.

Our work is concerned with the austenite to ferrite phase transition in iron, where we consider the inter-phase interface to be formed according to NW OR or some derivative of it. For this particular OR, the inter-phase boundary is classified as a semi-coherent interface, having an array of dislocations (termed as misfit dislocations), which help to partially reduce the misfit strain, originating from the lattice mismatch across the interface. It is now widely accepted that experimentally observed ORs do not correspond exactly to NW (or KS). The deviations can appear in the form of ledges or disconnections, which exhibit both step character and dislocation properties and higher index habit planes are observed at the inter-phase boundary.\cite{Shiflet1994, shiflet} 

Mainly because of importance of steel as a structural material, austenite to ferrite phase transition in iron is a widely researched topic. Various aspects of $\alpha-\gamma$ transformation have been investigated in pure iron\cite{LEE20137399,bos, tateyama, PhysRevB.80.214108,Sandoval_APL, song1, song2, song3, Song_2015,Tripathi_2018,urbassek,SINCLAIR20084160}, iron alloys \cite{WANG2014399_urassek_Fe_C, Wang_2014_Fe_C_MSME}, thin films\cite{Wang_2013_thiFilm}, single crystals \cite{sietsma_acta18,Karewarcryst9020099}, bicrystals \cite{karewar2020atomistic}, and nanowires\cite{Sadndoval_NL,Sandoval_2009}. Several studies, based on molecular dynamics (MD) simulations, cover kinetics and atomic mechanisms during the transformation, for flat\cite{bos, urbassek, Sandoval_2009}, as well as ledged interfaces \cite{song1,song2,Tripathi_2018}. Bos et al. \cite{bos} performed MD simulations for several possible ORs and concluded that martensitic transformation takes place with the aid of glissile screw dislocation networks nucleated at the interface during the incubation time. Tateyama et al.\cite{tateyama} found a planar and needle-like growth in case of NW and KS OR, respectively. They also reported a decrease in interface velocity when the parent and product phase was rotated in the range of $0-5.26^\circ$ from NW to KS OR. Wang and Urbassek \cite{urbassek} conducted MD simulation to uncover the pressure and temperature dependence of the FCC-BCC transformation using Meyer-Entel potential and identified the phonon softening, leading to the phase transition. Song and Hoyt \cite{song2} investigated an inter-phase boundary obeying NW OR, along with structural disconnections (sessile in nature) and reported the nucleation and movement of a secondary set of glissile disconnections at the terrace plane, leading to the FCC-BCC transformation. Ou et al. \cite{ou_w} investigated three semicoherent interfaces joined by NW, KS, and Nagano ORs and found martensitic transformations in the coherent areas of the interface (having low potential energy), while some diffusional jumps were also observed in the incoherent areas (having high potential energy). Karewar et al. \cite{sietsma_acta18,Karewarcryst9020099} conducted a MD simulation of martensitic transformation in single-crystal Fe and Fe-C systems with different planar defects such as twin boundaries and stacking faults (SFs). They observed several well-known transformation mechanisms (such as NW, KS, Burgers path \cite{BURGERS1934561}, and BB/OC model \cite{BOGERS1964255, OLSON1972107}) depending on the type of the defect structure present in the simulation system. Tripathi et al. \cite{Tripathi_2018} reported that the ledges or disconnections at the interface are the preferential nucleation sites for the BCC phase and interface movement was guided by the growth of the BCC phase along the ledge vector. Maresca et al. \cite{maresca_acta18} also studied the motion of the ledged FCC/BCC interface. A recent MD study by Karewar et al. \cite{karewar2020atomistic} discussed the role of misfit dislocations on the martensitic transformation in case of a flat interface. The authors observed that the screw dislocations within the interface plane govern the atomic displacements, leading to the phase  transformation.

The phenomenological theory of martensitic transformation \cite{BHADESHIA20141021} provides the orientation relationship between the parent and product phases but does not elaborate on the atomistic displacements during the transformation. Similarly, most of the MD studies mentioned in the previous paragraph\cite{bos, tateyama, PhysRevB.80.214108,Sandoval_APL, song1, song2, song3, Song_2015,Tripathi_2018,urbassek,SINCLAIR20084160,WANG2014399_urassek_Fe_C, Wang_2014_Fe_C_MSME, Wang_2013_thiFilm, Sadndoval_NL,Sandoval_2009, urbassek, karewar2020atomistic,maresca_acta18,Sadndoval_NL,ou_w} do not explicitly  discuss the movements of the individual or group of atoms during the transformation. Further, all the studies focus on transformation in the presence of either flat or ledged interfaces, lacking a clear qualitative or quantitative comparison among different interface morphologies. Moreover, there exist conflicting reports on the mobility of the interface, with some studies claiming the flat interface to be immobile,\cite{song1,song2,song3,Tripathi_2018} contradicting with others.\cite{bos,tateyama,ou_w} In order to bridge this gap, one needs to compare transformation in flat vs. ledged interfaces using exactly similar simulation parameters like boundary conditions, interatomic potential, thermostat and barostat etc. In this work, we attempt to resolve this issue by studying atomic mechanisms during the austenite to ferrite phase transformation in iron using classical MD simulations. We capture several \textit{on-the-fly} pictures of atomic displacements during the martensitic transformation. By doing this, we are able to provide a comprehensive description of the martensitic transformation in iron; specifically, the effect of morphology, by comparing flat vs. ledged interfaces, which has not been done so far (to the best of our knowledge). We identify two distinct types of atomic motions leading to the phase transformation; one set of movements (mainly along the misfit dislocation network present at the interface) are confined to the interface planes and the other having out-of-the interface plane component as well (the latter observed exclusively in case of the ledged interface). Interestingly, the temperature dependence of the interface mobility is found to show opposite trends in case of flat and ledged interfaces.

\section{Simulation Details}
\label{sd}

Atomic trajectories are obtained using classical MD simulations, as implemented in Large-scale Atomic/Molecular Massively Parallel Simulator (LAMMPS).\cite{lammps} An embedded atom method (EAM) based potential, developed by Ackland et al.\cite{ackland}, is used to describe the inter-atomic interactions among the iron atoms. Several materials characteristics, like lattice parameter, cohesive energy, vacancy formation energy, and elastic constants, predicted by this potential are known to be in good agreement with the experimental data, as well as density functional theory (DFT) based predictions.\cite{Tripathi_2018} However, this empirical potential overestimates the melting point of Fe\cite{sunprb2004} and also fails to correctly predict the austenite-ferrite transition temperature of 1185 K. Despite these limitations, Ackland potential has been used in several MD studies of iron, including the FCC-to-BCC phase transformation.\cite{song1, song2, ack_ref, ack_ref2}    

\begin{figure}
\includegraphics[width=1\linewidth]{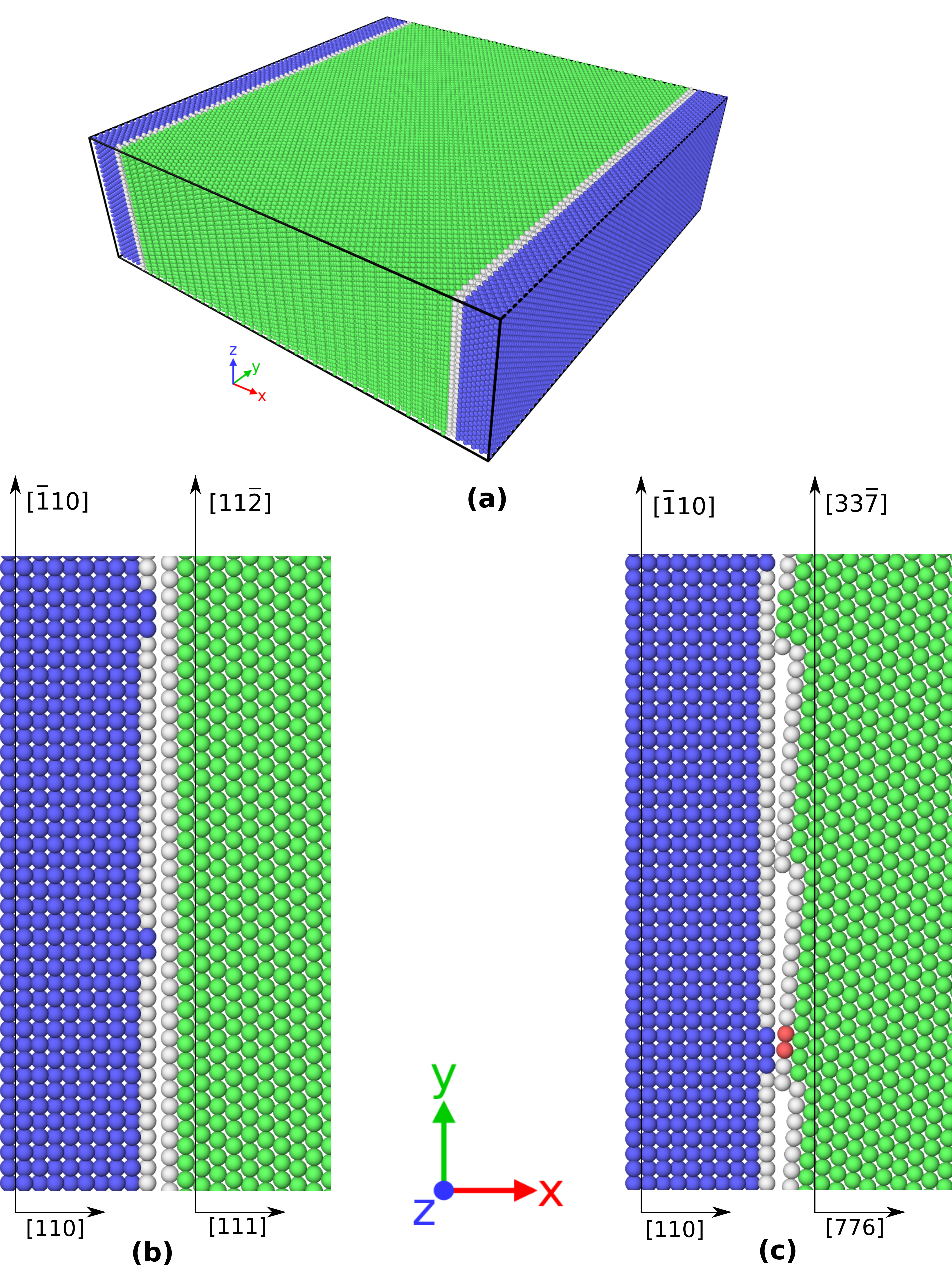}
\caption{(a) Sandwich structure of BCC-FCC-BCC simulation system, used in this study, (b) Close up view near the flat interface. (c) Close up view at the ledged interface showing the atomic steps or disconnections. The atoms are color-coded as per a-CNA; blue: body centered cubc (BCC), green: face centered cubic (FCC), red: hexagonal close packed (HCP), and grey: unidentified atoms, not belonging to any of the three. Same atomic color coding is followed in the rest of the figures.}
\label{fig1}
\end{figure}

 \begin{table}
\caption{Details of crystallographic orientations parallel to the $x$, $y$, and $z$ directions of the simulation boxes used in this work.}
\begin{tabular}{|c|c|c|c|c|c|}
\hline
Phase & Direction &   Orientation   & \makecell{Dimensions\\ (\AA)} & \makecell{Tilt\\ angle} & \makecell{No. of\\ atoms} \\
\hline
\makecell{BCC \\ (Flat)}  & \makecell{x \\ y \\ z}  & \makecell{ $[110]$ \\ $[\bar{1}10]$ \\ $[001]$} & \makecell{20.39 \\ 260.97 \\  77.85} & - & 69120 \\
\hline
\makecell{BCC \\ (Ledged)}  & \makecell{x \\ y \\ z}  & \makecell{ $[110]$ \\ $[\bar{1}10]$ \\ $[001]$} & \makecell{20.39 \\ 240.58 \\  77.85} & - & 63720 \\
\hline
\makecell{FCC\\ (Flat)}   & \makecell{x \\ y \\ z}  & \makecell{ $[111]$ \\ $[11\bar{2}]$ \\ $[\bar{1}10$]} & \makecell{261.17 \\ 261.71 \\ 78.15} & 0$^\circ$ & 351480\\
\hline
\makecell{FCC\\ (Ledged)} &  \makecell{x \\ y \\ z}  & \makecell{ $[776]$ \\ $[33\bar{7}]$ \\ $[\bar{1}10$]} & \makecell{213.24 \\ 241.25 \\ 78.15} & 4.04$^\circ$ & 321600 \\
\hline
\end{tabular}
\label{T1}
\end{table}

A sandwich structure of the BCC-FCC-BCC sequence is created, such that the $x$-axis is normal to the phase boundary, which is aligned parallel to the $yz$ plane [Figure~\ref{fig1}(a)]. The crystallographic orientation of $x$, $y$, and $z$ direction of the BCC and FCC regions, and their dimensions (at T=600 K) are listed in Table \ref{T1}. After creating the two regions separately, they are joined together to get the final simulation box of BCC-FCC-BCC sandwich shape [Fig.~\ref{fig1}(a)]. Spacing between BCC and FCC phase is taken as an average of inter-planar spacing of both the phases. The $y$ and $z$ dimensions of the BCC and FCC phases are chosen very carefully, such that the mismatch of the dimensions at the $yz$ cross-section is minimal (less than 0.5\%). A large mismatch leads to high stresses at the phase boundary, which may affect the transformation behavior. The sandwich shaped simulation box remains periodic and without any free surfaces in each direction. 
  
Two types of $\gamma-\alpha$ interfaces, namely flat  [Figure~\ref{fig1} (b)] and ledged [Figure~\ref{fig1} (c)], are created. The flat interface is created using ideal NW OR, such that \newline \centerline{$\{111\}_{fcc} || \{110\}_{bcc} ; [\bar{1}10]_{fcc} || [001]_{bcc}$.} In case of the ledged interface, the FCC phase is rotated about the $z$-direction (parallel to the $[1\bar{1}0]$ direction of FCC) by an angle of $4.04^\circ$, with respect to the ideal NW OR [Table \ref{T1}]. As a result of this, $x$ and $y$ axis becomes parallel to the crystallographic directions $[776]$ and $[33\bar{7}]$, respectively, in the FCC phase. Because of the rotation, the ledged interface contain steps with one atom height in the FCC side of the interface [Fig.~\ref{fig1}(c)], similar to the vicinal surface. A detailed analysis of the phase boundary is given in our previous work \cite{Tripathi_2018}, while the important features relevant for the current work are shown in Fig S1 in the Supporting Information. According to adaptive common neighbor analysis (a-CNA)\cite{Stukowski_2012_aCNA}, most of the atoms present at the phase boundary does not belong to the parent BCC or product FCC phase, but marked as some unidentified category, which happens because of the lattice parameter mismatch between the parent phases. 
After creating the sandwich shaped simulation box, both BCC and FCC phases are individually equilibrated, first using NVT ensemble for 100 ps to bring all the atoms in thermal equilibrium. Subsequently, NP$_x$T ensemble is used for volume equilibration for another 100 ps, keeping $y$ and $z$ dimensions fixed. This reduced the stresses in the simulation system to less than $\pm$20 MPa. After equilibrating the BCC and FCC phases independently, phase transformation dynamics are executed using the NPT ensemble for 5-30 ns (depending on interface type and temperature). During the phase transformation, each dimension was allowed to equilibrate independently to reduce the transformation stresses.  
\begin{figure}
\includegraphics[width=1\linewidth]{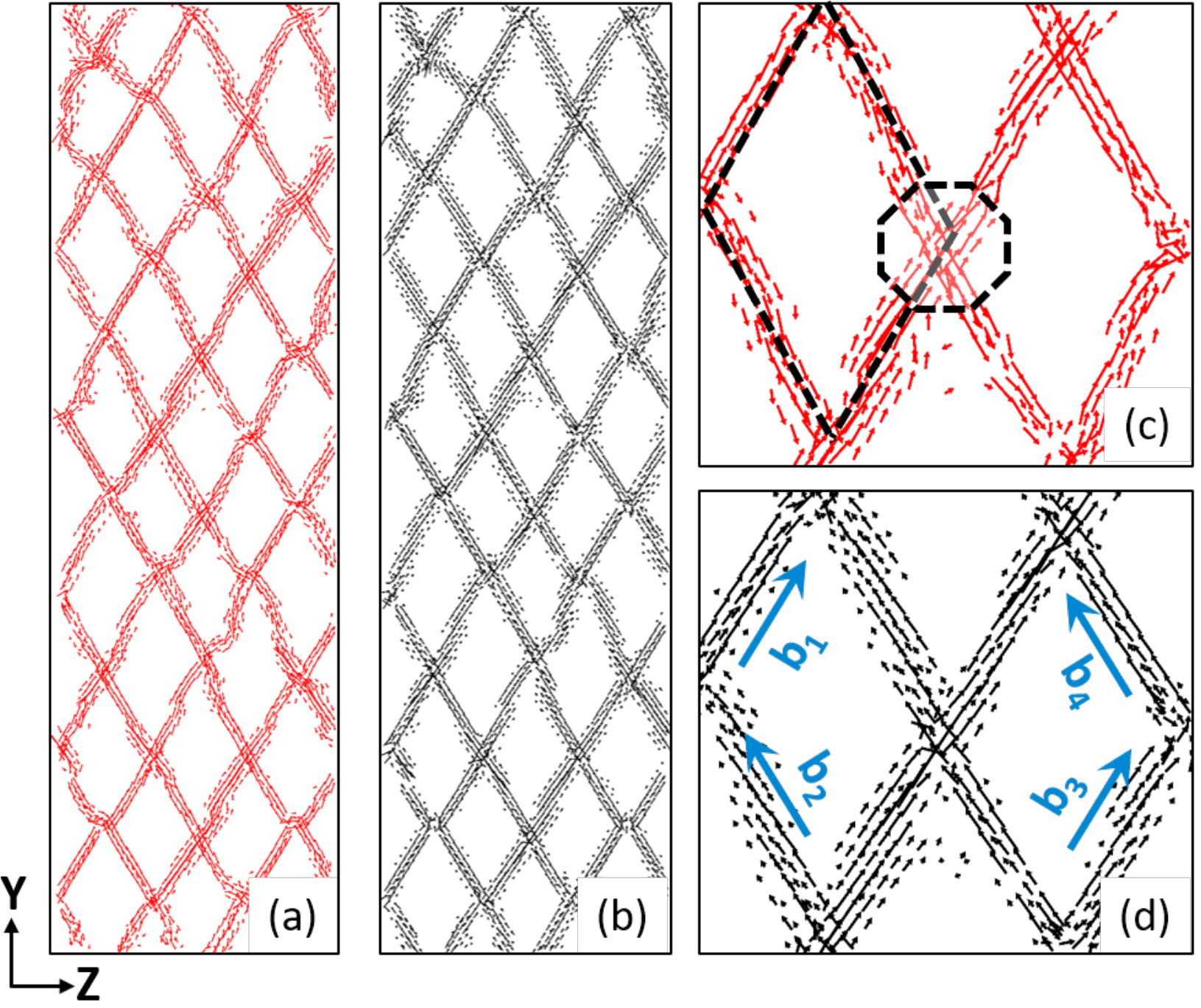}
\caption{The misfit dislocation structure at the flat interface analyzed by NTA at 100 K. The image is taken after the entire simulation system is equilibrated and before the transformation starts. (a) Misfit dislocation line directions (red arrows), (b) Burgers vectors (black arrows), and their enlarged views are shown in (c,d). Incoherent regions of the interface are formed along the dashed lines (making a diamond-like pattern) and particularly, within the octagon. Atoms in these regions are having high potential energy, as shown in Figure S2 of the Supporting Information.}
\label{fig:flat_NTA}
\end{figure}
 
Four different values of temperature (600 K, 800 K, 1000 K, and 1200 K) are chosen to study the transformation kinetics in the flat and ledged interface. The thermal vibrations of the atoms at high temperatures make it difficult to identify the atomic displacements responsible for FCC-to-BCC transformation. Therefore, atomic positions averaged over 1 ps are used for the purpose of visualization and data analysis. The atomic displacements are further scaled by a factor of 1.75 for better visibility. Detailed analysis of the interface (in terms of misfit dislocations) is carried out at an even low temperature of 100 K, than that of transformation dynamics (600 to 1200 K). We use Nye Tensor analysis in the assigned mode (NTA),\cite{HARTLEY_Nye,NYE1953153} as implemented in AADIS (atomistic analyzer for dislocation character and distribution) code, \cite{AADIScode} to study the misfit dislocation structure at the interfaces. The analysis helps us to characterize the dislocation line directions and Burgers vectors of the misfit dislocations. 

\begin{figure*}
\includegraphics[width=0.9\linewidth]{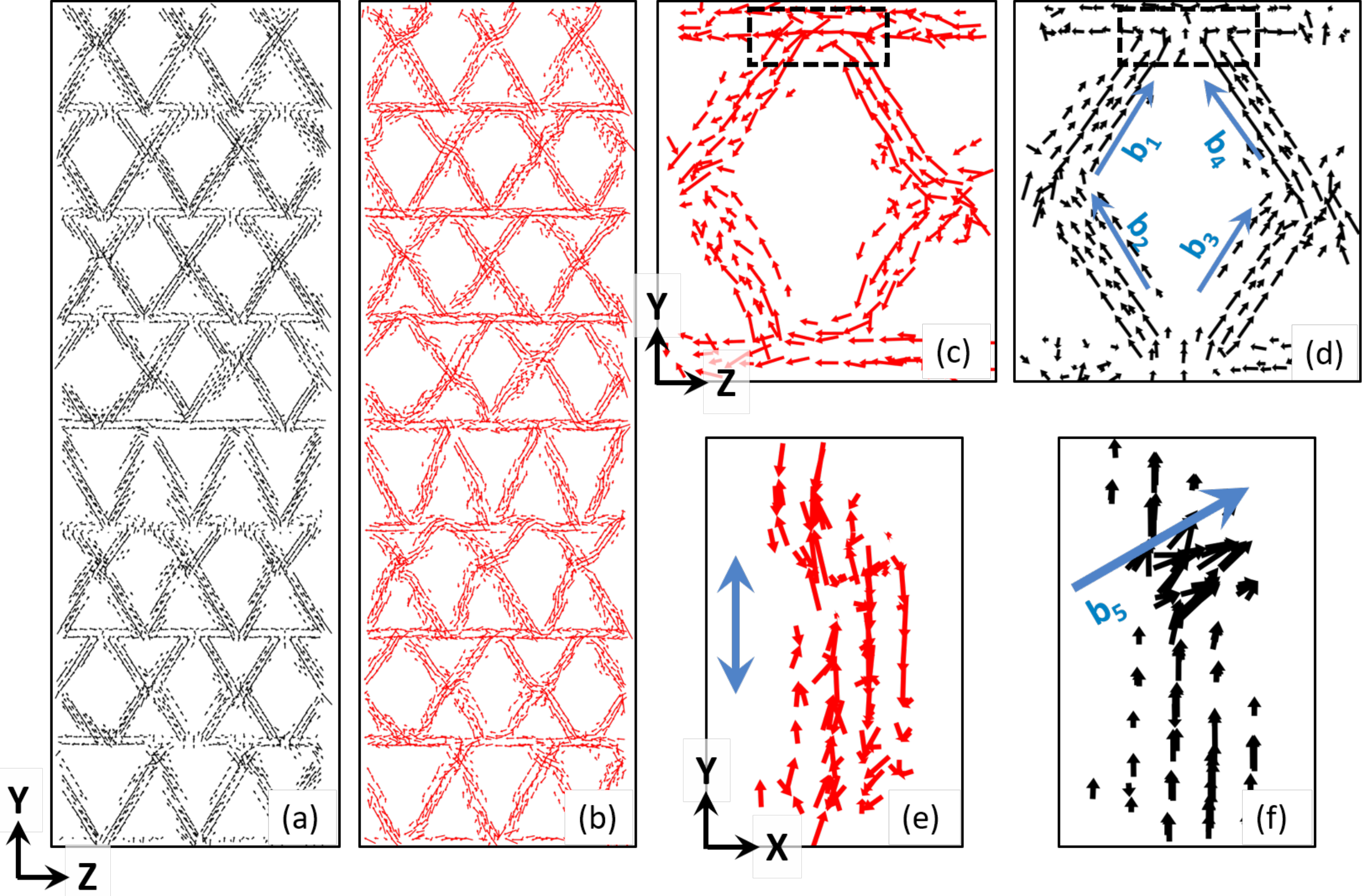}
\caption{The misfit dislocation structure at the ledged interface analyzed by NTA. (a) Dislocation line directions, (b) Burgers vector directions, and a close-up view (c-d) in YZ plane. (e-f) A further close-up view (in XY plane and within the rectangular area marked in (c-d)) of one of the ledges; also showing some out-of-the interface plane component of the Burger vector $b_5$. Such out-of-the interface plane components are not observed in the case of the flat interface.}
\label{fig:ledged_NTA}
\end{figure*}

\section{Results}
\label{rd}

\subsection{Structure of the interfaces}
Before studying the transformation, first we analyze the atomic structure of the interface using NTA, which reveals the presence of misfit dislocations at the BCC-FCC interface. In case of the flat interface, the dislocation line directions (red line) and the Burgers vectors (black line) are separately shown in Figures  \ref{fig:flat_NTA}(a) and (b), respectively. Interestingly, the dislocation network forms a diamond-shaped pattern, where the dislocations are lined along the edges of the diamond, while the area inside forms the coherent region of the interface. For a better visibility, a close-up view of the interface is shown in Figures \ref{fig:flat_NTA}(c-d). As seen in this figure, the dislocation line directions and the Burgers vectors are parallel to each other, which indicates that the misfit dislocations present at the interface are screw dislocations, with Burgers vectors along $b = <111>_{bcc}$. The dislocation line directions and Burgers vectors are parallel to the interface plane ($yz$ plane), and there is no component out-of-the interface plane. Such a screw dislocation network at the FCC/BCC interface has previously been postulated by Howe et al.\cite{hirth_review} Further analysis reveals that the flat interface consists of areas with low and high potential energy, as shown in Figure S2 in the Supporting Information. Comparing Figure \ref{fig:flat_NTA} with Figure S2, it can be seen that low potential energy areas lie within the diamond, showing a coherent FCC-BCC interface in these regions. On the other hand, areas with misfit dislocations (along the edges of the diamond and in particular, the octagon marked in Figure \ref{fig:flat_NTA}) have higher potential energy, resulting from incoherent stacking of the FCC and BCC phase.  

Figure \ref{fig:ledged_NTA}(a-b) shows misfit dislocation network structure at the ledged interface. Similar to the case of the flat inter-phase boundary, the diamond-shaped network of screw dislocation is formed at the ledged interface also. However, due to the presence of the ledges, the pattern is not as regular, as observed in the case of the flat interface. Figure \ref{fig:ledged_NTA}(c-f) presents the enlarged view of the NTA at one of the diamond regions for better visibility, clearly showing that the dislocation line directions along the edges of the diamond are parallel to the interface ($yz$ plane), having no out-of-the interface plane component.

However, an important distinction (compared to the flat boundary) arises near the ledges, where an additional set of dislocations appear, with Burgers vectors also having an out-of-the interface plane component. As shown in figure \ref{fig:ledged_NTA}(d,f), unlike the dislocation lines, Burgers vector $b_5 = <110>_{fcc}$ on the $\{111\}$ plane has noticeable $x$-component. The SFs are nucleated from this region when the transformation starts (to be discussed later in subsection \ref{sect_transform_ledged}). Atoms located near the ledges are found to have the highest potential energy, as shown in Figure S3, Supporting Information. 

\begin{figure}
\includegraphics[width=1\linewidth]{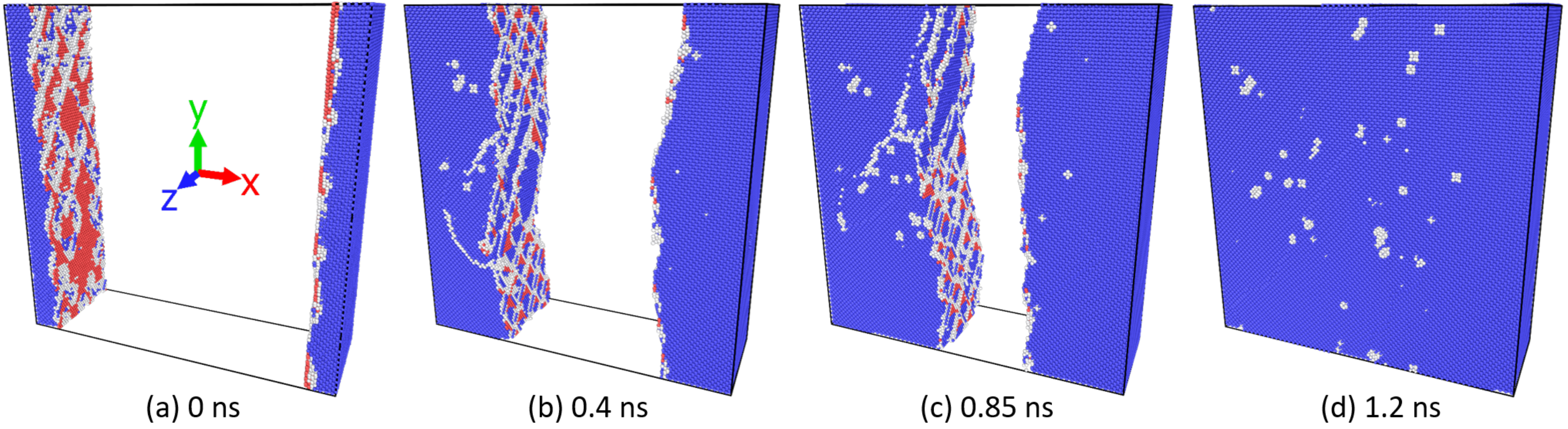}
\caption{The FCC-to-BCC transformation at 600 K for the simulation system with the flat interface. Atoms with FCC crystal structure are not shown for better visibility. (a) snapshot after equilibration at 0 ns; various stages during the transformation at (b) 0.4 ns, (c) 0.85 ns; and after the transformation is over at (d) 1.2 ns.}
\label{flat_Xmation_overview}
\end{figure}

\begin{figure*}
\includegraphics[width=1\linewidth]{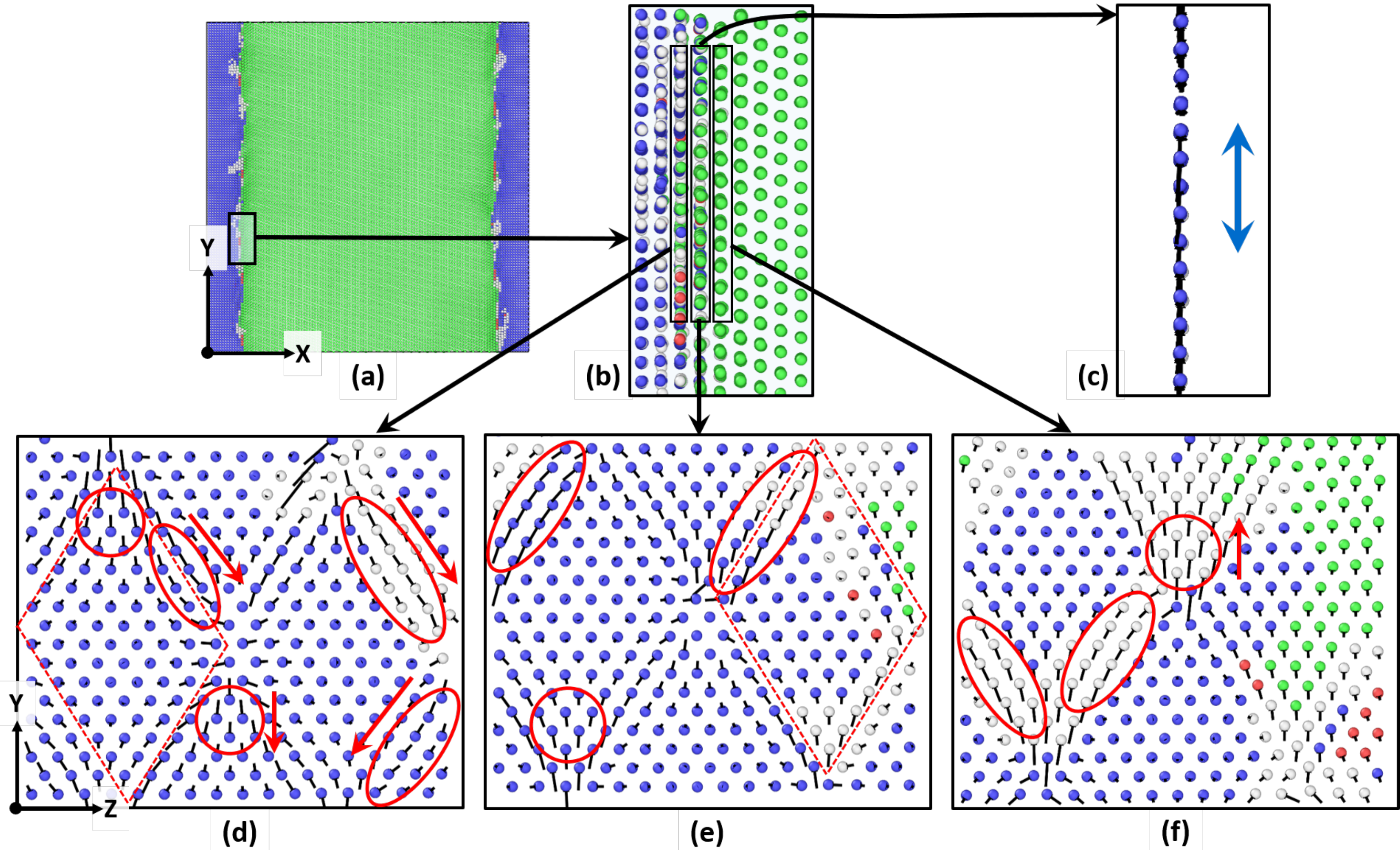}
\caption{Atomic displacements observed during the FCC-to-BCC transformation in the presence of a flat interface at 600 K. (a-b) Snapshots at 0 ps, just after equilibration and before the transformation starts. Snapshots at 10 ps after the transformation starts, illustrated for the (c) $XY$ plane and (d-f) $YZ$ plane, showing displacements causing the phase transformation to be mainly localized in the $YZ$ plane. The red dashed-diamonds are guide to the eye for the original positions of the misfit dislocation networks.}
\label{flat_atomic_Xmation}
\end{figure*}

\subsection{Transformation mechanisms: Flat interface}
Figure \ref{flat_Xmation_overview} shows the overview of the transformation for the flat interface at 600 K. The structure at 0 ns (just after equilibration) is shown in figure \ref{flat_Xmation_overview}(a), which is the starting configuration for the transformation simulations. Two intermediate states at 0.4 and 0.85 ns are shown in figures \ref{flat_Xmation_overview}(b-c) and the transformation is completed at 1.2 ns [Figure \ref{flat_Xmation_overview}(d)]. From the figures, it is clear that the transformation from the FCC-to-BCC phase happens only at the interface. Otherwise, the BCC phase would have nucleated somewhere inside the FCC phase, which is not observed in any of our simulations.

Since the transformation happens at the phase boundary, we further investigate the atomic displacements (leading to the FCC-to-BCC transformation) near the interface. Figure \ref{flat_atomic_Xmation}(a) shows the configuration just after equilibration (0 ns), and a zoomed-in view of the area marked by the rectangle is shown in subfigure (b). Atomic displacements at one of the FCC monolayers (at the interface) is captured during the transformation at 10 ps [Figure \ref{flat_atomic_Xmation}(c)]. As shown in the figure, all the atomic displacements are parallel to the $YZ$ plane (blue arrow showing the atomic movements vertically upwards or downwards), having no out-of-the interface plane component.  

Having established that the atomic displacements responsible for the phase transition are confined to the interface planes, we now analyze atomic motions in these planes. The zoomed-in view of three monolayers (originally FCC $\{111 \}$ planes) adjacent to the interface are illustrated in figures \ref{flat_atomic_Xmation}(d-f) in $YZ$ projection (parallel to the interface). Coordinated movements of several groups of atoms (confined to the interface planes) can be observed in all three monolayers. Moreover, the atomic displacements are found to take place mainly in the regions, where the misfit dislocation networks were located in the beginning. The same type of atomic displacements observed in the diamond region [Figure \ref{flat_atomic_Xmation}(d-f)] are also found on the entire monolayer at the interface, as well as few adjacent layers in the FCC region. The atomic displacements at the interface plane create an atomic drag, that propels the atoms on the neighboring atomic layers to move in a similar fashion. This interface governed transformation propagates through the entire FCC phase layer-by-layer until the transition is complete, as shown in the Supplementary Information Figure S4.

In the case of a flat interface, the transformation is aided by the screw dislocations at the interface. These screw dislocations (having high PE) move on the $\{111\} <110>_{fcc}$ or $\{110\} <111>_{bcc}$ slip system, as shown by red arrows and red ellipses in Figure~\ref{flat_atomic_Xmation}(d-f), thereby reducing the PE and facilitating the phase transformation. This mechanism is the same as the Kurdjmov-Sachs (KS) FCC-to-BCC transformation mechanism \cite{Kurdjumow1930}. The corresponding FCC-to-BCC orientation relationship during the transformation can be written as,
\centerline{$(111)_{fcc} || (110)_{bcc}, [10\bar{1}]_{fcc} || [11\bar{1}]_{bcc}$} 
\centerline{$(111)_{fcc} || (110)_{bcc}, [\bar{1}0\bar{1}]_{fcc} || [\bar{1}1\bar{1}]_{bcc}$} \newline 
\centerline{$(111)_{fcc} || (110)_{bcc}, [101]_{fcc} || [111]_{bcc}$} \newline
\centerline{$(111)_{fcc} || (110)_{bcc}, [\bar{1}01]_{fcc} || [\bar{1}11]_{bcc}$.}

On the other hand, within the red circles, atomic movements as per the Burgers-Bogers-Olson-Cohen (BB/OC) method \cite{BOGERS1964255,OLSON1972107} are observed [Figure~\ref{flat_atomic_Xmation}(d-f)]. As per the BB/OC method, the BCC phase nucleates as a result of the two shears, $T/3$ and $3T/8$, where $T = a/6<112>$ is the Burgers vector of the Shockley partial for the FCC twin shear. During the transformation, regions of HCP stacking is observed, as shown in the Supplementary Information Figure S2 and S4, which is the meta-stable structure after the first atomic shear $T/3$. The atoms in this meta-stable state move further according to the second shear ($3T/8$), completing the  transformation. Figure S5 in the supplementary material shows the schematic of the BB/OC mechanism and similar transformation pathways are observed in the present simulations. The KS and BB/OC mechanisms were also reported in previous MD studies of phase transformations in single crystal and bicrystal iron \cite{karewar2020atomistic,ou_w,sietsma_acta18,Karewarcryst9020099}. 

\begin{figure}
\includegraphics[width=1\linewidth]{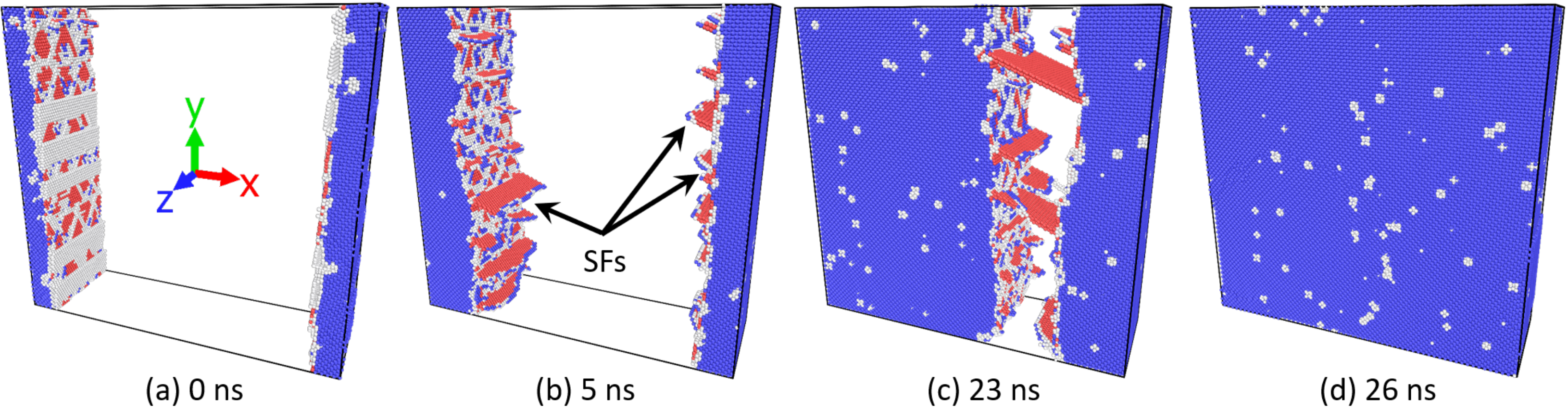}
\caption{The FCC-to-BCC transformation at 600 K for the ledged interface. Atoms with FCC crystal structure are not shown for better visibility. (a) snapshot after equilibration at 0 ns; various stages during the transformation at (b) 5 ns, (c) 23 ns; and after the transformation is over at (d) 26 ns.}
\label{ledged_Xmation_overview}
\end{figure}

\begin{figure*}
\includegraphics[width=0.8\linewidth]{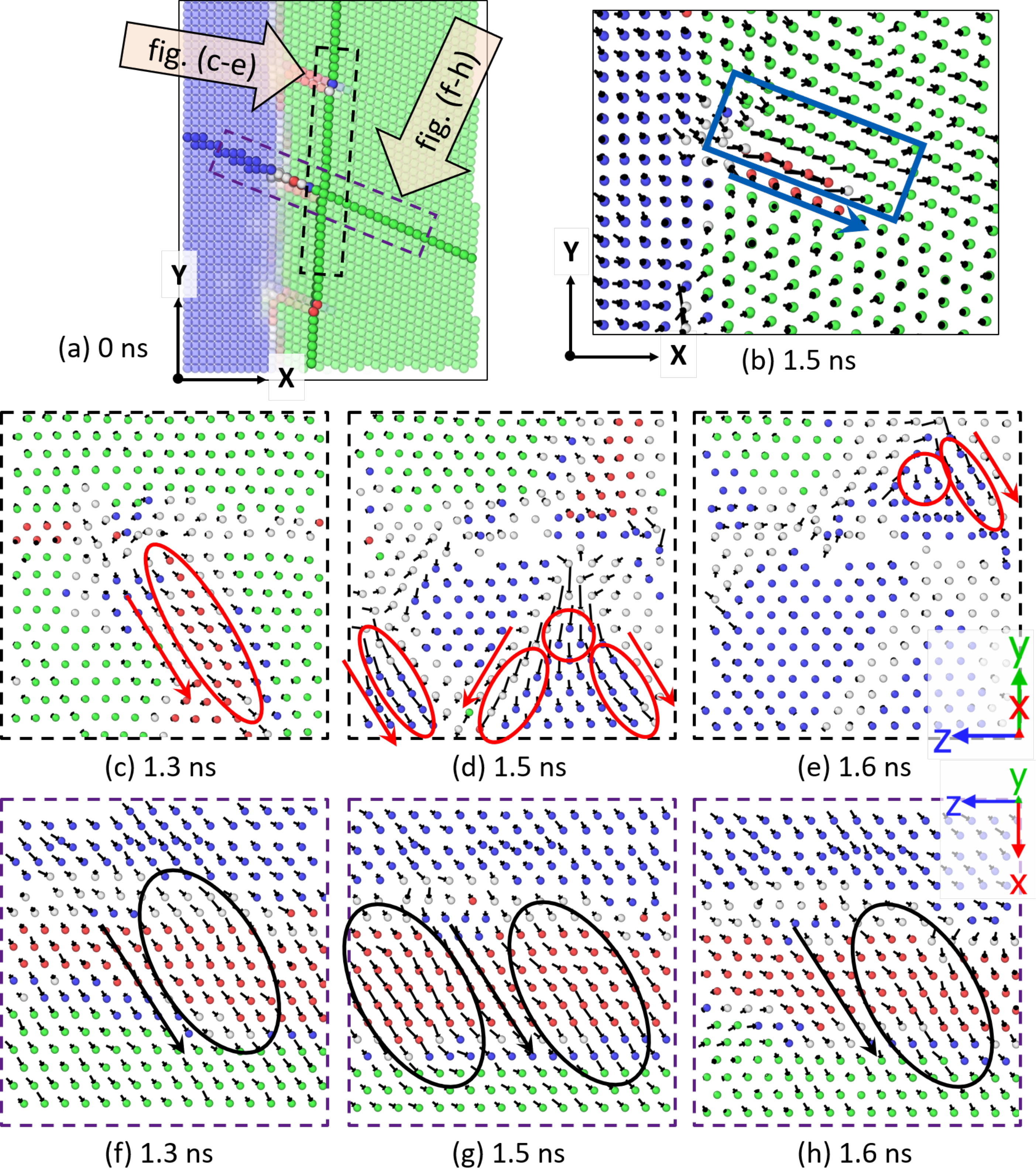}
\caption{Atomic displacements observed during the FCC-to-BCC transformation in the presence of ledged interface at 600 K. Few layers close to the interface are shown at (a) 0 ns, before the transformation starts and (b) 1.5 ns, in $XY$ projection. The close-up view of the highlighted monolayers are shown in (c-e) $YZ$ and (f-h) $XZ$ projection.}
\label{ledged_atomic_Xmation}
\end{figure*}

\subsection{Transformation mechanisms: Ledged interface}
\label{sect_transform_ledged}

Figure \ref{ledged_Xmation_overview}(a-d) shows the evolution of the martensitic transformation in the presence of the ledged interface at 600 K. The equilibrated structure of the simulation box at 0 ns is shown in figure \ref{ledged_Xmation_overview}(a). Two intermediate states are shown at 5 ns and 23 ns [Figures \ref{ledged_Xmation_overview}(b-c)] and the transformation is completed at 26 ns [Figure \ref{ledged_Xmation_overview}(d)]. In addition to the transformation mechanisms found at the flat interface, an additional feature is observed in case of ledged inter-phase boundary. During the process of transformation, stacking faults (SFs) are nucleated from the atomic ledges [Figure \ref{ledged_Xmation_overview}(b-c)]. The nucleation of SFs at the atomic ledges has been observed previously in pure Fe.\cite{song1,song2, Tripathi_2018} These SFs play an important role in FCC-to-BCC phase transformation, which is going to be described in detail in the following discussion. 

The atomic displacements inducing the FCC-to-BCC transformation are shown in Figure \ref{ledged_atomic_Xmation}. The close-up view of the interface at 0 ns (just before starting the transformation), which includes few layers from both FCC and BCC phase, is shown in Figure \ref{ledged_atomic_Xmation}(a). One can clearly see the atoms with HCP stacking near the ledges, where SFs are nucleated. Figure \ref{ledged_atomic_Xmation}(b) shows the close-up view of the atomic displacements near the interface on a monolayer, in the same $XY$ projection plane as in sub-figure (a). The interface in this figure is perpendicular to the $x$ direction; and thus, it can be concluded that, in the presence of the ledged inter-phase boundary, out-of-the interface plane atomic displacements (marked by the blue rectangle) also takes place, in addition to the in-plane atomic displacements (similar to the case of the flat interface planes), leading to the phase transformation.

To analyze the in-plane and out-of-the interface plane atomic displacements in ledged interfaces, we investigate the two highlighted monolayers (within the dashed rectangles), as shown in Figure \ref{ledged_atomic_Xmation}(a). For these monolayers, the atomic displacements (in a plane perpendicular to the arrows) are shown in sub-figures (c-e) and (f-h). 
Sub-figures \ref{ledged_atomic_Xmation}(c-e) show the atomic displacements on $\{111\}_{fcc}$ plane. The atomic displacements in this monolayer lie within the interface plane and there is no out-of-plane component.  Similar to  the flat interface [Figure \ref{flat_atomic_Xmation}(d-f)], the atomic displacements on this monolayer [Figure\ref{ledged_atomic_Xmation}(c-e)] follow two mechanisms- (i) KS mechanism shown in the red ellipses and (ii) BB/OC mechanism shown within the red circles.


Sub-figures \ref{ledged_atomic_Xmation}(f-h) show the monolayers parallel to the stacking fault nucleated from the ledges. The atomic displacements on this layer follow the HCP-to-BCC Burgers path on the $\{1\bar{1}00\}$ $<11\bar{2}0>_{hcp}$ slip system shown within the black ellipses. Figure S6 [Supporting Information] further shows the similarity between the atomic displacements as per the Burgers path and the displacements observed in the present simulations. These type of out-of-the interface plane displacements are mainly observed on the atomic layers at and near the SFs. Therefore, the SFs, nucleated from the ledges affect the martensitic transformation by generating an additional out-of-the interface plane atomic displacement component. 

\section{Discussion}
\label{dis}

\begin{table}
\caption{Gibbs free energy change (Eq. 1 in Supporting Information) and interface velocities calculated (Eq. 2 in the Supporting Information) at different orientations and temperatures. These numbers are calculated by taking an average of the values obtained from eight independent simulations starting with different initial velocities for each of the temperature and orientation.}
\begin{tabular}{|c|c|c|c|}
\hline
\makecell{ Tempera- \\ture (K)}& \makecell{  $\Delta(G)_{\gamma-\alpha}$\\ (kJ/mole)} & \makecell{ Flat interface \\ velocity (m/s)} & \makecell{ Ledged interface \\ velocity (m/s)} \\
 \hline
\makecell{ 600} &\makecell{2.20} & \makecell{ 13.46$\pm$0.95 }& \makecell{0.21$\pm$0.05  }\\
 \hline
 \makecell{800  } & \makecell{2.06  } & \makecell{10.33$\pm$0.48  }& \makecell{0.78$\pm$0.04  }  \\
 \hline
 \makecell{1000} & \makecell{1.91} &  \makecell{ 7.66$\pm$0.31 } & \makecell{1.37$\pm$0.06  }  \\
 \hline
 \makecell{1200} & \makecell{1.76} & \makecell{ 5.69$\pm$1.23 } & \makecell{2.06$\pm$0.32 } \\
 \hline
\end{tabular}
\label{Table:DeltaG_intf_vel}
\end{table}

As seen in the previous section, in the case of the flat interface, FCC-to-BCC transformation takes place by only in-plane (of the interface) atomic shear component. On the other hand, an additional set of out-of-the interface plane displacements are observed in the presence of the ledged interface. The two types of displacements in presence of the ledged interface (in-plane and out-of interface plane) compete with each other during the transformation. The out-of-plane atomic displacements caused by the SFs create a barrier to the in-plane atomic movements (due to the screw dislocations) and thereby slow down the transformation rate in systems having ledged interfaces, compared to the ones with flat interfaces.

\begin{figure}
\includegraphics[width=\linewidth]{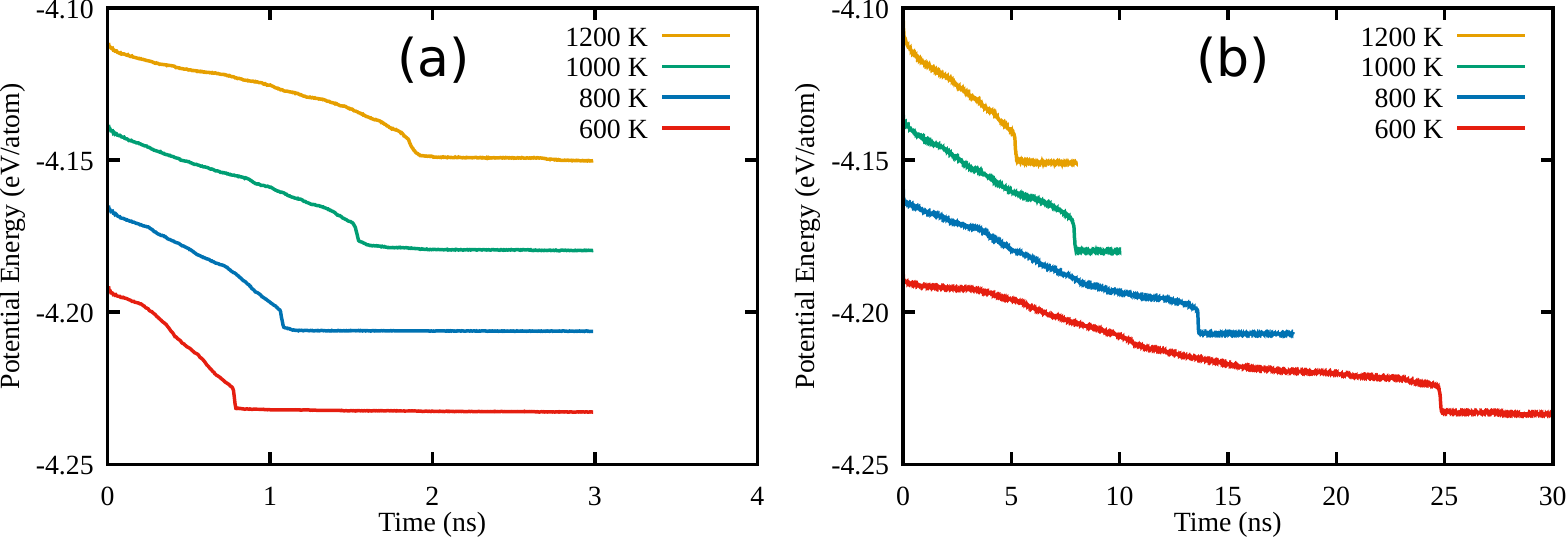}
\caption{Potential energy profile of atoms in (a) flat NW, and (b) rotated NW i.e. ledged interface as a function of time. Lines become horizontal once the transformations are over.} 
\label{fig:Potene}
\end{figure}

\begin{figure}
\includegraphics[width=1\linewidth]{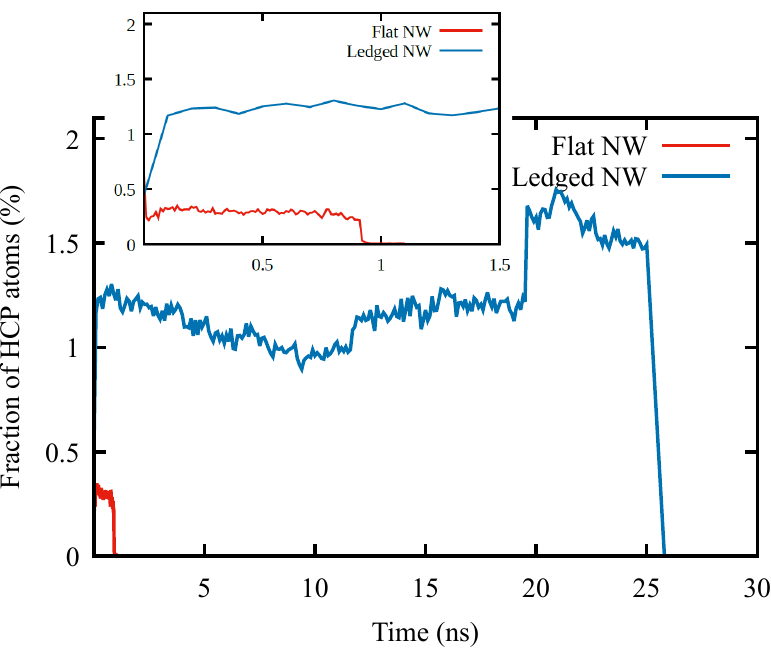}
\caption{The fraction of atoms with HCP stacking during transformation for the flat and ledged  interfaces at 600 K. The inset shows an enlarged view near the origin of the graph.}
\label{fig:hcpatoms}
\end{figure}

In figure \ref{fig:Potene}, it is also observed that for the systems with flat interfaces, the transformation rate increase with a decrease in temperature. On the contrary, for systems with the ledged interface, the transformation slows down with decreasing temperature. Such contrasting behavior in the presence of the two different types of interfaces needs thorough analysis and is beyond the scope of the present work. Our preliminary analysis suggests that for the flat interfaces, the transformation is mainly driven by the reduction of the Gibbs free energy $\Delta(G)_{\gamma - \alpha}$. Using a similar approach used previously by Tripathi et al. \cite{Tripathi_2018}, thermodynamic integration at different temperatures are carried out to get the values of $\Delta(G)_{\gamma - \alpha}$ at different temperatures, as shown in table \ref{Table:DeltaG_intf_vel}. It can be seen that the value of $\Delta(G)_{\gamma - \alpha}$ increases with decreasing temperature. The increase in the driving force at lower temperatures suggests faster kinetics in the case of systems with flat interfaces.

The transformation in the systems with the ledged interface is mainly controlled by the defect mobility i.e. SF movement in the simulation system. While the calculation of the SF mobility at different temperatures is beyond the scope of this paper, our results suggest that the mobility of the SFs affects the velocity of the interface. Lower velocity of the SFs results a lower velocity of the interface and thereby a slower transformation kinetics and vice versa. The interface velocities are reported in the table \ref{Table:DeltaG_intf_vel}, calculated using the approach previously mentioned in the literature \cite{song1,Tripathi_2018}. The velocity of the ledged interface increases with an increase in temperature from 600 K to 1200 K, which is possibly due to the higher SF mobility at a higher temperature. 

Since the presence of SFs significantly impacts the transformation rate, further analysis is carried out in terms of the fraction of SFs (or fraction of atoms with HCP stacking) present in the simulation systems. The fraction of atoms with HCP stacking in systems with different morphologies are compared in Figure \ref{fig:hcpatoms} at 600 K. As seen here the fraction of atoms with HCP stacking is much higher in the presence of a ledged interface than that of a flat interface. A similar trend is observed at higher temperatures as well. This is because of the nucleation and propagation of the SFs from the atomic ledges. Moreover, the fraction of atoms having HCP stacking remains almost unchanged during transformation in systems with the flat interface; on the other hand, it increases during transformation (at 20 ns) in the presence of a ledged interface in the system. This increase in the fraction of SFs happens when the two interfaces, from left and right, come closer and the SFs propagating from them merge [Figure \ref{ledged_Xmation_overview}(c)]. As expected, no atoms with HCP stacking are found once the FCC-to-BCC transformation is completed.

\section{Conclusions}
\label{con}

In summary, using classical molecular dynamics simulations and an embedded atom method type interatomic potential, we illustrate the effect of the morphology of the FCC/BCC interface on the martensitic transformation in iron. We compare two types of morphologies: a flat one, where the phases are joined according to NW OR and a ledged one, where the FCC phase is rotated by $4.04^\circ$ with respect to the ideal NW OR. In case of the flat morphology, we find misfit screw dislocations, having their Burger vector component lying within the interface plane. We identify the atomic displacements, parallel to the interface plane and taking place along the misfit dislocation network, leading to the phase transition, following the KS and BB/OC FCC-to-BCC transformation mechanism. In addition to the in-plane (parallel to the interface) movements, we also find out-of-the interface plane motion, contributing to the phase transition in case of ledged interface. We find some SFs to be nucleated at the ledges, causing the atomic shear as per the HCP-to-BCC Burgers path, leading to transformation from the parent FCC to product BCC phase, via the intermediate HCP phase (appearing as SFs). We find the mobility of the inter-phase boundary to be lower in case of the ledged interface, as SFs hinder the interface motion. Interestingly, the transformation kinetics shows opposite trend as a function of temperature in case of the flat and ledged interface. In presence of the flat interfaces, the kinetics is controlled by the driving force for the transformation (i.e. $\Delta G_{\gamma - \alpha}$). In presence of the ledged interface, the mobility of the defects (SFs) controls the velocity of the interfaces and which subsequently controls the transformation kinetics. We believe that our study will provide a better understanding of austenite to ferrite phase transformation in iron and help to bridge the existing gaps in the literature.


\section{Acknowledgements}
S. B. acknowledges funding from SERB (CRG/2019/006961). Authors acknowledge HPC IITK and NCHC NCTU Taiwan for providing computational facilities.

Author contribution: P. K. T. and S. K. contributed equally to this work.
\bibliography{ref}

\end{document}